\documentclass[pra,aps,twocolumn,groupedaddress,superscriptaddress,floatfix]{revtex4}
\usepackage{epsfig}
\usepackage{graphicx}
\usepackage{amsfonts}
\usepackage{amsmath}

\normalsize

\begin{document}


\title{Correlated two-photon imaging with true thermal light}
\author{Da Zhang$^1$, Xi-Hao Chen$^{1,2}$, Yan-Hua Zhai$^1$, and Ling-An Wu}
\affiliation{Institute of Physics, Chinese Academy of Sciences, Beijing 100080 \\
$^2$ Department of Physics, Liaoning University, Shenyang 110036,
China}

\begin{abstract}
We report the first experimental demonstration of two-photon
correlated imaging with true thermal light from a hollow cathode
lamp. The coherence time of the source is much shorter than that
of previous experiments using random scattered light from a laser.
A two-pinhole mask was used as object, and the corresponding thin
lens equation was well satisfied. Since thermal light sources are
easier to obtain and measure than entangled light it is
conceivable that they may be used in special imaging applications.
\end{abstract}

\maketitle

Although imaging is an old and well-studied topic and is of great
importance in classical optics, it is now attracting new interest
in quantum optics due to recent experiments on two-photon
correlated imaging. The first such experiment was based on quantum
entangled photon pairs from spontaneous parametric down conversion
in a nonlinear crystal, and gave rise to the name ``ghost''
imaging, so called because an object in one optical beam could
produce an image in the coincident  counts with a detector placed
in another beam~\cite{Pittman}. This experiment led to other
interesting theoretical and experimental studies  and,
furthermore, a debate on the question whether entanglement is a
prerequisite for ghost imaging~\cite{Abouraddy}. The possibility
to perform correlated imaging with thermal light was first
predicted by Gatti et al~\cite{Gatti2}. The first experiment with
a classical light source that demonstrated ``two photon''
coincidence imaging was performed by Bennink et al. using a
coherent laser beam split along two paths with detectors that
measured finite laser pulses~\cite{Bennink1}.

The difference between quantum and classical coincidence imaging
and the extent to which a classical light source can mimic a
quantum one have been widely discussed by the groups of Shih
\cite{Shih,Scarcelli}, Boyd~\cite{Bennink2},
Lugiato~\cite{Gatti2,Gatti1,Gatti3}, Zhu~\cite{Zhu,Yangjian} and
Wang~\cite{Cao}. Experimentally, Shih and collaborators first
achieved ghost imaging with a pseudo-thermal light source, and
introduced the concepts of ``two-photon coherence'' and
``two-photon incoherence'' imaging~\cite{Alejandra}. Gatti et al.
obtained high resolution ghost imaging with thermal-like speckle
light. However, in all these experiments the primary light source
was a He-Ne laser, and the pseudo-thermal beam was obtained by
passage through a rotating ground glass plate~\cite{Gatti4}.

Different from these experiments, we report the demonstration of a
two-photon correlated imaging experiment using a true thermal
light source.

We employed a commercial rubidium hollow-cathode
lamp~\cite{Bernhard} manufactured by the General Research
Institute for Nonferrous Metals (China), which is the type
commonly used in atomic absorption spectroscopy because of its
sharp spectral linewidth. The lamp was powered by a direct current
of 20mA in our experiments, and the resonance wavelength was
780nm. However, the actual linewidth of hollow-cathode lamps
depends on the pressure, filament structure etc and varies from
model to model. In our model the inner diameter of the cathode was
3mm.
\begin{figure} \label{fig1}
\includegraphics[width=8cm]{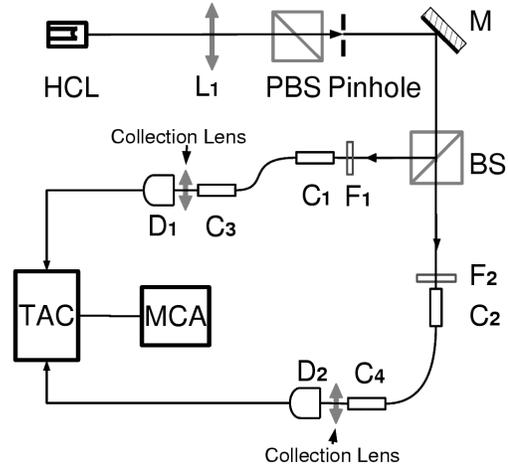}
\caption{Schematic of the HBT type experiment. HCL: hollow cathode
lamp; $L_1$: lens of focal length 10cm; pinhole diameter: 0.5mm;
effective diameter of collimators is 2mm.}
\end{figure}
\begin{figure} \label{fig2}
\includegraphics[width=7cm]{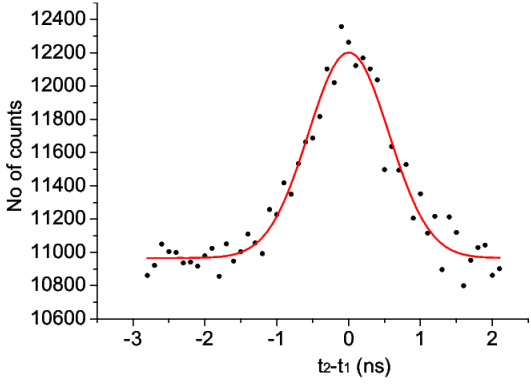}
\caption{Number of counts vs time interval for the HBT type
experiment. The solid curve is a Gaussian fit of data points. The
FWHM of the peak is about 1.3ns which is due to the time jitter of
the electronic circuits.}
\end{figure}
To estimate the coherence time of our lamp we first carried out a
Hanbury Brown-Twiss (HBT) type experiment, with the setup shown
schematically in Fig.~1. The light from the lamp is focused by the
convex lens ($L_1$) of 10cm focal length onto a circular pinhole
0.5mm in diameter to form a secondary light source. A polarizing
beam splitter (PBS) in front of the pinhole transmits linearly
polarized light. After reflection by a mirror (M) the beam is
divided by a $50\%/50\%$ non-polarizing beam splitter (BS). The
reflected and transmitted beams pass through interference filters
$F_1$ and $F_2$ before being coupled into single photon detectors
$D_1$ and $D_2$, respectively, through fiber collimators $C_1$ and
$C_2$. The transmission of the interference filters is about
$70\%$ at 780nm and the receiving area of the collimators is about
2mm in diameter. The detector output signals are sent to a
time-amplitude converter (TAC), with $D_1$ and $D_2$ providing the
``start'' and ``stop'' signals, respectively. The TAC output is
connected to a multi-channel analyzer (MCA), and the computer
displays a histogram of the different intervals between the times
of arrival of the photons at the two detectors. From this we
obtain the relation between the photon count rate and time
interval, and subsequently the second-order correlation function
\begin{eqnarray*}  \label{G2}
 G^{(2)}(t_2-t_1) = \langle \hat{E}_{2}(t_2)^{(-)}
 \hat{E}_{1}(t_1)^{(-)}
 \hat{E}_{1}(t_1)^{(+)} \hat{E}_{2}(t_2)^{(+)}  \rangle,
\end{eqnarray*}
here
$\hat{E}_{i}^{(-)}(t_i)$ are the
positive and negative frequency field operators at detectors $D_i
(i=1,2)$ at time $t_i$, respectively. The normalized second-order
correlation function $g^{(2)}(t_2-t_1)$, which describes the
intensity correlation of a light field at two
detectors~\cite{Loudon}, is given by
\begin{eqnarray}  \label{g2}
g^{(2)}(t_2-t_1) &=&
\frac{G^{(2)}(t_2-t_1)}{G^{(1)}_{1}(t_1)G^{(1)}_{1}(t_2)},
\end{eqnarray}

If the average intensity of the light remains constant~\cite{Kolobov},
\begin{eqnarray*}
\lim_{(t_2-t_1) \to \infty} G^{(2)}(t_2-t_1) =I_1 I_2 = G^{(1)}_1(t_1)G^{(1)}_2(t_1).
\end{eqnarray*}
we obtain the value $g^{(2)}(t_2-t_1=0)$ experimentally by
dividing the values of the average $G^{(2)}(t_2-t_1)$ for values
of, $|t_2-t_1| \leq 0.25ns$ (the width of the coincidence window)
by the value of $G^{(2)}(t_2-t_1)$ for $|t_2-t_1| \gg 1.3ns$
(corresponding to signals arriving randomly beyond any correlation
time).

A typical set of data from our experiment is shown in Fig. 2. The
full width at half maximum (FWHM) of the peak is about 1.3ns,
which is much longer than the coherence time of the light, and is
mainly due to the time jitter in the electronic circuits. From the
above data we derive the coherence time $\tau_0$ to be about
0.2ns~\cite{hbtbook}, which is much shorter than that of previous
experiments using randomly scattered light from a He-Ne laser. The
maximum of the measured normalized second-order correlation
function $g^{(2)} $ is 1.11, corresponding to a maximum visibility
of about 5$\%$. The deviation of the measured $g^{(2)}$ from the
theoretical value 2 is due to the area of the non-point-like
collimator, as well as the time jitter of the electronic circuits.
When we increased the area of the light source (the size of the
pinhole), the value of $g^{(2)}$ decreased further, as expected.

\begin{figure} \label{fig3}
\includegraphics[width=8cm]{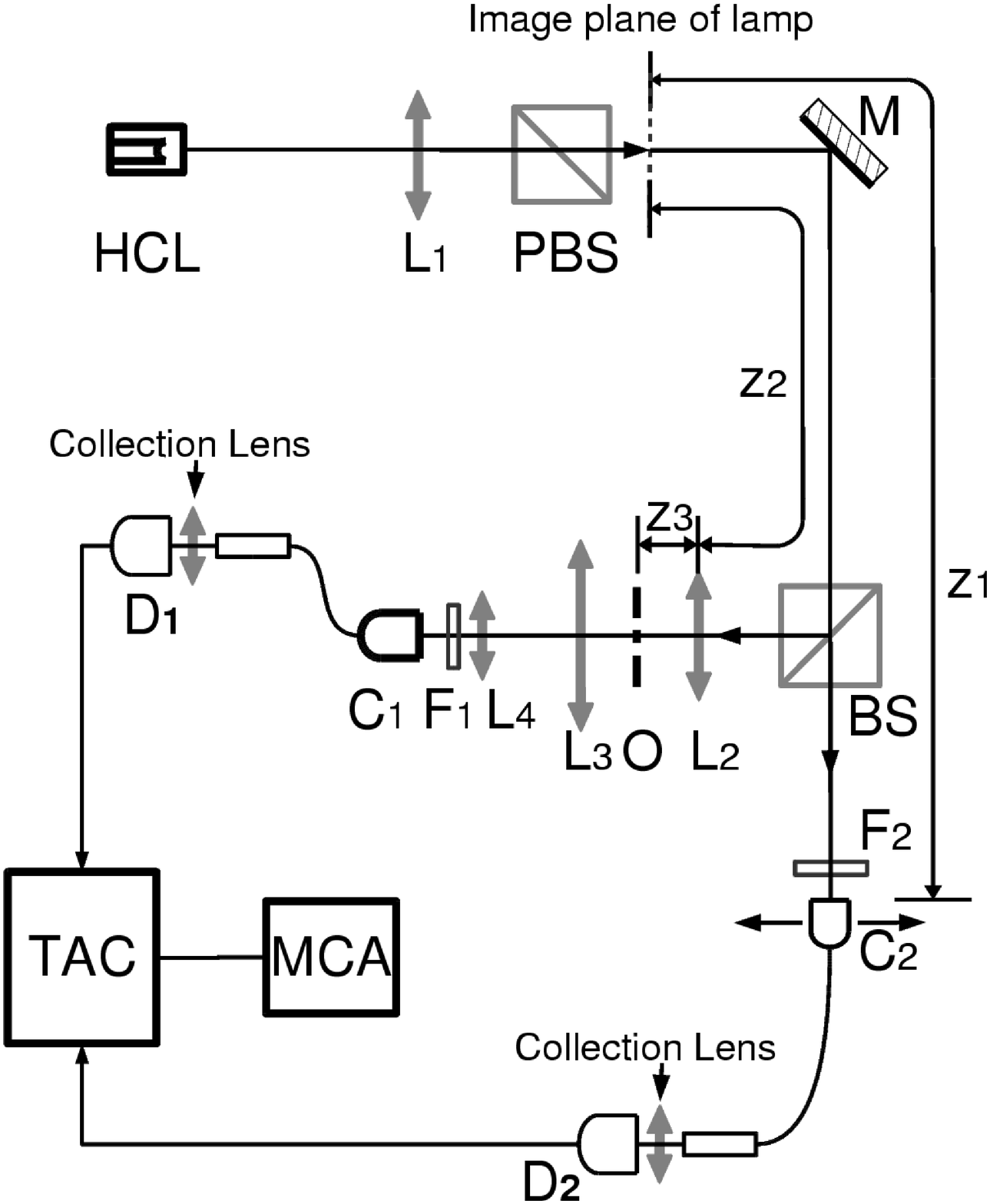}
\caption{Experimental set-up of the ghost imaging experiment.
Focal length of $L_2=20$cm, $z_1-z_2=32.5$cm, $z_3=12.4$cm,
$z_1=1.8m$; the image of the cathode $\approx$ 1mm in diameter.}
\end{figure}


For our correlated ghost imaging experiment the set-up in the HBT
type experiment is modified as shown in Fig. 3. The pinhole in the
image plane of the lamp is removed for greater light throughput.
The object, a mask (O) consisting of two pinholes,
is inserted in the beam reflected from the beamsplitter BS.
The diameter of both pinholes is 0.5mm and the distance between
them is 1.3mm.
Two lenses $L_3$ and $L_4$ act as a telescope so that $D_1$ can
capture all the light passing though the mask and serve as a
bucket detector.  The third lens $L_2$ inserted between the beam
splitter and the mask is the convex imaging lens, and has a focal
length of about 20cm. Note that this set-up is different from
those of references 4 and 6, and the corresponding Gaussian thin
lens equation is
\begin{eqnarray}  \label{gaussianlensequ}
\frac{1}{z_2-z_1}+\frac{1}{z_3} =\frac{1}{f},
\end{eqnarray}

\begin{figure} \label{fig5}
\includegraphics[width=8cm]{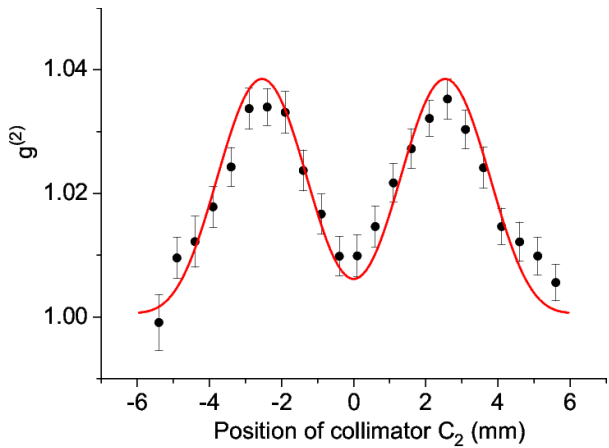}
\caption{Dependence of the normalized second order correlation
function $g^{(2)}$ on the position of fiber collimator $C_2$,
which gives the cross sectional image of the two pinholes. The
solid curve is calculated taking into consideration the finite
size of the detectors~\cite{hbt}.}
\end{figure}

where $z_1$ and $z_2$ are the distances from collimator $C_2$ and
lens $L_2$ to the secondary light source, respectively, and $z_3$
is the distance from lens $L_2$ to the mask O; $f$ is the focal
length of $L_2$.  This equation can be found in
references\cite{Yangjian,Cao} for incoherent ghost imaging, and is
the same as that for quantum ghost imaging~\cite{Pittman}, except
that $+z_1$ is replaced by $-z_1$.

The transverse normalized second-order correlation function is
given by~\cite{Alejandra}:
\begin{eqnarray} \label{g2 for ghost image}
g^{(2)}(x_2) \propto N + |T(\frac{z_3}{z_1-z_2}x_2)|^2,
\end{eqnarray}
where $x_2$ is the transverse position of fiber collimator $C_2$,
$T(x)$ the transmission function of the mask (O), and $N$ the
number of transparent features in the object plane, which equals 2
in our scheme because there are two pinholes in the mask. This
equation reflects the position-position correlation between the
object and image planes, as well as the fact that the visibility
decreases (background increases) when the number of points in the
object increases.

In our experiment we choose the case of the object distance
$z_3<f$ ~\cite{Cao}, with  $z_3=12.4$cm and image distance
$z_1-z_2=32.5$cm . The fiber collimator $C_2$ is scanned
transversely across the reference beam in steps of 0.5mm, and the
detector coincidence counts recorded as above for the HBT type
experiment. The actual single count rates of $D_1$ and $D_2$ were
about 300k/sec, and about 30k counts were accumulated for each
data point. The normalized second-order correlation function
$g^{(2)}$ was calculated as above, from which we plot the cross
sectional image of the two-pinhole object, as shown in Fig. 4. The
visibility is found to be $2\%$.

Apart from the factor $N$ in Eq. \ref{g2 for ghost image}, other
reasons for the lower visibility include the short coherence time
compared with the time jitter of the detection system, the limited
coherence area in the object plane, which is affected by the area
of light source, and the finite area of the fiber collimator
$C_2$. The low resolution is due to the finite area of scanning
fiber collimator and the finite coherence area in the object
plane, which is about $0.35mm^2$.


In conclusion, we have experimentally realized ghost imaging with
true thermally incoherent light, showing that  thermal light can
emulate the role of entangled light in ghost imaging experiments
to a certain extent. Although the visibility is very low this
could be greatly improved by removing the background by some
means, e.g. digitally. Since thermal light sources are easier to
obtain and measure it is conceivable that they could find certain
special applications~\cite{ShenshengHan} where entangled sources
are not so convenient to use.

We are grateful to Yan-Hua Shih, De-Zhong Cao, Kai-ge Wang, and
Shi-Yao Zhu for useful discussions. We also thank Zhan-Chun Zuo
and Hai-Qiang Ma for their experimental assistance. This work was
supported by the National Natural Science Foundation of China
(Grant No. 60178013) and the Knowledge Innovation Program of the
Chinese Academy of Sciences.

\end{document}